\documentclass[preprint,5p]{elsarticleGAC}

\usepackage{etoolbox}
\makeatletter
\patchcmd{\ps@pprintTitle}{\footnotesize\itshape
       Preprint submitted to \ifx\@journal\@empty Elsevier
       \else\@journal\fi\hfill\today}{\relax}{}{}
\makeatother

\usepackage{amsmath}
\usepackage{amsfonts}
\usepackage{amssymb}
\usepackage{times}
\usepackage{graphicx}

\begin{document}

\title{Phenomenology of Philosophy of Science: OPERA data}

\author{Giovanni AMELINO-CAMELIA\\
{\small{\it{Dipartimento di Fisica, Sapienza Universit\`a di Roma and INFN, Sez.~Roma1, P.le A. Moro 2, 00185 Roma, EU}}}}
%

\vskip -0.9cm

\begin{abstract}
\noindent
I observe that,
as the physics side of the OPERA-anomaly story is apparently unfolding,
there can still be motivation for philosophy of science to analyze
the six months of madness physicists spent chasing the dream of a new
fundamental-physics revolution.
 I here mainly report data on studies of the OPERA anomaly
  that could be relevant for analyses from the
  perspective of phenomenology of philosophy of science.
  Most of what I report is an insider's perspective on
   the debate that evolved from the original announcement
  by the OPERA collaboration of evidence of superluminal neutrinos.
  I also sketch out, from a broader perspective,
  some of the objectives I view as achievable for
  the phenomenology  of philosophy of science.
\end{abstract}

\maketitle

\baselineskip11pt plus .5pt minus .5pt

Part of the work I do for my day job, as a theoretical physicist,
is aimed at establishing a "quantum-gravity phenomenology"~\cite{dawn}.
Until not long ago the expression "quantum-gravity phenomenology" was thought
to be an oxymoron (see, {\it e.g.}, Ref.~\cite{isham}).
But it is now a real scientific programme,
or at least so it is thought to be by a still small but fast-growing number of physicists
(examples of quantum-gravity-phenomenology studies, strongly biased
toward my own works, are in
Refs.~[3-12]).

I do realize that what I am here proposing may appear (and perhaps is)
even more nonsensical: I will argue that at least a certain part of philosophy
of science (the one that revolves around the notion of ``theory of knowledge")
constitutes a science in itself.
And of course by this I mean that those aspects of philosophy
of science are not well served confining their analysis to the realm
of endless debates. Instead they can and should be subjected to an appropriate form of
experimental scrutiny.

From my limited perspective
as amateur philosopher it appears that this is already done to some extent.
Certain "data" are used when philosophy contemplates
questions such as "what is knowledge?" (or~\cite{stachelWHERE} ``where is knowledge?"),
 "what is the scientific method?",
"does knowledge grow continuously or only through periods of scientific revolution?".
The data which are used are data of the same nature I am here reporting,
but they are typically taken from the history of science (often from the
history of fundamental physics, which is also the context I am here focusing on).
So at least naively one can view philosophy of science
as a science based on postdictions. Perhaps this is why most scientists
remain unimpressed by the philosophy-of-science debate: it comes with the
trade that we pay no attention to postdictions.

Lacking the needed training in philosophy for articulating my case
logically, I am here going to use\footnote{{\bf An earlier version of these notes was proposed
to Nature last February. A one-page extract
(essentially focusing on the possible implications for the ``blind-analysis standards",
which is one of the issues I shall consider
here) was then published as Nature 483 (2012) 125.}} as illustration of the perspective I advocate
the context of the recent debate on the superluminal-neutrino
anomaly reported by the OPERA collaboration~\cite{operaOLD,opera}.

For six months it felt, at least to a theoretical physicist such as myself,
as if everyone was intrigued by the OPERA anomaly: even physicists working
outside fundamental physics, even other scientists, even the media,
even those working at the small grocery store on the ground floor of my building.\\
But I could see no trace of philosophers of science. I must have missed
some interest which surely developed in the right exclusive small circles of philosophers.
But what
I imagined would happen from the philosophy side had to be of such a magnitude
I could not possibly miss it.\\
 Where were all those theorists of scientific revolutions?
How hard were they trying to guess whether this was the dawn of a new scientific
revolution? Or was this "normal science"?\\
Which of the tens of models proposed for the OPERA anomaly went into
a degenerative phase more quickly? which ones experienced at least a 48-hour window of
progressiveness? If the OPERA anomaly survived a bit longer which of those
tens of research programmes should have been more generously funded?\\
And, if OPERA had not found a systematic error, would special relativity
be {\underline{falsified}}?\\
I imagined the philosophy echo would be a bang I could not miss.
It must have been a blip which I did miss.


\section{OPERA's philosophy-of-science data}

\subsection{Preliminaries: asking ``what did we learn?"}
Before actually going to the philosophy-of-science data reporting,
I want to comment on the many paths that can take
 theoretical physics
to new ``knowledge".
I will do this  borrowing from my own familiarity,
as an interested onlooker~\cite{gacbjorken},
with the interest generated by the
so-called "centauro events"~\cite{centauro}:
some cosmic-ray events were being reported with highly asymmetric
structure (like a centauro) and producing nearly exclusively charged pions,
whereas our current
particle-physics theories suggest that in such cases typically one third of the pions should
be neutral. Motivated in part by the desire of
explaining the centauro events the community figured out that something roughly
(but, and I stress this, not exactly) similar to a centauro event is actually predicted
by our theories in certain peculiar circumstances. "Inspired" by the centauro events
these previously unnoticed and potentially important aspects of particle physics
(particle-physics theory)
  were understood. And yet it is now rather well established~\cite{centauro}
that the centauro events were an artifact of
the limitations of the relevant experiments.

So my readers should keep in mind that
I am adopting a rather broad perspective on what could constitute knowledge.
And I tend to assume that a variety of paths can lead to such knowledge.
But I do not think we can proceed as it appears to be suggested by some philosophers,
who choose to confine their analysis of the connection
between ``method" and "knowledge" to
the appreciation of the fact that the ideal path to discoveries/knowledge would be
one of unlimited resources for all methodologies. Even within my amatorial philosophy one can
easily appreciate the fact that ideally this would maximize our knowledge. But the way science
actually works
is strongly affected by the fact that our resources are far from unlimited. The limited resources available
 impose  difficult choices for assigning different priority among different methodologies
 and even among different research programmes
adopting essentially the same methodology. Part of my motivation for proposing a phenomenology of philosophy of
science comes from the conviction that philosophy of science cannot shy away from these challenges involved
in assigning priority and funding. The science of philosophy of science should, in my view, play a crucial
role in such decisions. Daring to assign more priority to one proposal over another, while making sure that
the overall distribution of resources keeps some balance, especially by not penalizing too severely
proposals on new methodologies. Those new methodologies would appear less promising than methodologies with
an already established knowledge/discovery track record, and should therefore be assigned lower priority.
But some new methodologies
should receive at least modest encouragement, since we cannot exclude the possibility of new methodologies
eventually proving to be better (or better suited to new times) than previously successful ones.
This fits perfectly with the view of phenomenology of science as a science: rather than "statically best"
 theories of knowledge we should conceptualize, in the sense of
Lakatos,  "research programmes" in phenomenology of science, and expect that those research programmes
will experience stages of superiority over competing programmes followed by degenerative stages.

\subsection{21st-century scientific method}
A first few observations are warranted by the OPERA-related philosophy-of-science data
contributing to the impression that
something is changing in the way science is made. At least in fundamental physics
some macroscopic changes are in progress, probably rendering obsolete
(or perhaps urgently modern?) all previous discussions on the scientific method.
This is tangible to anyone whose work in fundamental physics,
as in my case, spans at least over the last 20 years. Fundamental physics
is different now (or at least works differently now) with respect
to the situation 20 years ago.

Part of this is visible in the time sequence of events related to the OPERA anomaly.
On September 22nd (2011) we had the first release~\cite{operaOLD}
of OPERA data on neutrino speeds,
but insiders were reached by rumors about it as early as a few weeks before,
and it was on every blog on the internet planet by a week before the announcement.
Then on November 17th (again insiders and blogs had known it for a few days)
there was a second announcement of
OPERA data on neutrino speeds~\cite{opera}.
The relation between the first and second announcement is of interest for
this manuscript in several ways. But
let me first note down that the two announcements were consistent with each other
and would imply speeds $\mathsf{v}$ such that
 $(\mathsf{v} - c)/c \simeq 2 \cdot 10^{-5}$
 (denoting as usual with $c$ the speed-of-light scale)
for muon neutrinos of a few
tens of GeVs
produced at Geneva's CERN laboratory and detected
at Gran Sasso's LNGS  laboratory (after a journey of about 730 Km through the Earth's crust).
A result of superluminal speed ($\mathsf{v} > c$) with a
nominal confidence of about six standard deviations~\cite{operaOLD,opera}.

Within some 48 hours of the first announcement, last september, of the OPERA findings~\cite{opera}
there were already a few analyses (see, {\it e.g.}, Refs.~[18-20])
pointing out that such superluminal speeds for $\sim 20 GeV$ neutrinos were not only
at odds with the principles of Einstein's relativity, but actually appeared to make little sense
also  when
compared with other determinations of neutrino speeds.
It was noticed~[18-20]
 that using neutrinos of energies about 3 orders of magnitude smaller
  observed from supernova SN1987a one can establish
a very conservative limit
on anomalous neutrino speeds at the level of 1 part in $10^9$.
And for neutrinos with energies of about an order of magnitude higher than
OPERA's we could rely~\cite{whataboutopera}
on the laboratory results reported in Ref.~\cite{preminos1979},
excluding peculiarities of neutrino speeds with accuracy
better than 1 part in $10^4$.
This led to the conclusion summarized in Fig.1: trusting OPERA
required not only giving up on Einstein relativity but also to do so
by introducing some rather fine-tuned new law for the speed
of particles.

\begin{figure}[htbp!]
\includegraphics[width=0.9 \columnwidth]{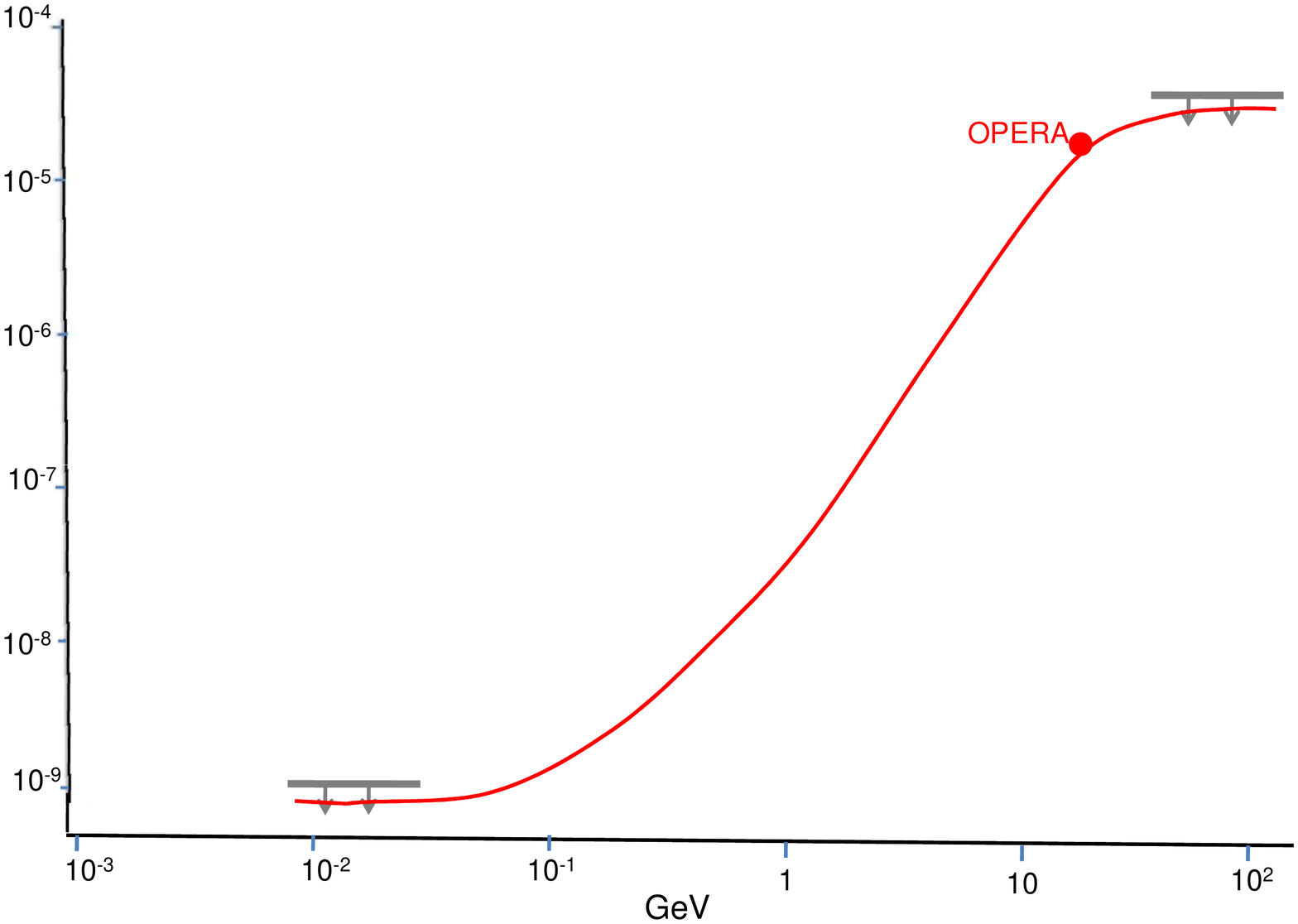}
\caption{A ``cartoonist impression" of how the energy dependence of neutrino speeds should have looked,
if we trusted the OPERA findings. The prediction of Einstein relativity over
all this range of energies is $\mathsf{v} - c \simeq 0$, since already at $10^{-3}$ GeV neutrinos
are effectively massless.
The OPERA result at $\sim 20 GeV$ should have been matched to the previous upper bounds
established around $\sim 10^{-2} GeV$, based on neutrino observations from SN1987a,
and the upper bound obtained with the travel-time studies for neutrinos of $\sim 100 GeV$
reported in Ref.~\cite{preminos1979}.}
\end{figure}

These considerations amplified the skepticism toward the OPERA result.
All the main aspects of the OPERA setup came under scrutiny in an unprecedented
example of what 21st century scientific method might look like, especially
exploiting the empowerment of internet communication.
Several concerns were raised~\cite{wrong1}
with respect to some of the GPS aspects of OPERA's
analysis and possible subtle roles played by
ordinary relativistic effects,
but these concerns were very quickly addressed~\cite{gpsOPERAnote,sagnacOPERAnote},
finding that they could not explain what OPERA had reported.

The objections which appeared to pose the most severe challenge
concerned (see, {\it e.g.},
  Ref.~\cite{bigpulseISbad})
  the statistical methods used in deriving
the  travel time estimated by OPERA.
That first version  of the OPERA analysis~\cite{operaOLD}
used data obtained with a setup such that the neutrino pulses produced at CERN
lasted about $10^4$ nanoseconds, and sophisticated inferences where needed to
derive the travel-time anomaly reported by OPERA, which is actually worth
 only $\sim 60$ nanoseconds. The statistical analysis also had to take into account
 that each neutrino in the pulses only has a tiny chance of interacting with the
  OPERA detector, since neutrinos interact only through the ``weak interactions".
  There was a (relatively) easy way to address such concerns:
  redo the experiment with neutrino
  pulses of shorter duration.
  I am sure I am not the only theorist to be impressed about how experimentalists
at CERN and LNGS managed to arrange exactly this new experiment in just a few weeks.
For the second run of the OPERA travel-time experiment, which is the main upgrade between
OPERA's first announcement~\cite{operaOLD} and its second announcement~\cite{opera},
 the neutrino pulses produced at CERN lasted $\sim 3$ nanoseconds.
 And the results of this second run are perfectly consistent
 with the previously announced results of the first run of the OPERA travel-time experiment.
 So we learned (we acquired knowledge about the fact)
 that the statistical analysis developed by OPERA in order to handle
 the time broadness of the original (first-run) neutrino pulses
 is a reliable way to perform such studies, which may prove valuable for future experiments.
 And a new, internet-based, way for scrutinizing puzzling experimental results
 was tried for the first time, proving to be of amazing effectiveness.
 The learning curve of anyone involved in studies of the OPERA anomaly was impressive,
 thanks to the fact that a world-wide network of scientists was kept well connected by the internet,
 and also exploiting the new rapid-publication methods afforded by internet, such as the Arxiv.

Overall there were more than 200
OPERA-inspired studies which appeared on arXiv over the six months
of debate.
This should be compared with the typical evolution of the
literature in most areas of fundamental physics:
for example in quantum-gravity research, a pretty active area,
one would say that a result has taken by storm the community
if it inspired 200 studies over a time of 4 or 5 years.
And it is important to notice that the six months it took for
this 200 OPERA-inspired studies are comparable to the standards for the time needed
by the refereeing process for a single paper to complete; indeed, this was all based on
the arXiv preprint by the OPERA collaboration.
In this respect
I should incidentally observe that, at least within fundamental physics,
the peer-review system had being showing signs of
obsoleteness increasingly over the last few years, and these aspects of the
OPERA story may suggest we should declare it officially dead.
The progress of the next fundamental-physics revolution
will not be recorded in conventional journals.

\subsection{Indirect bounds: OPERA had to be wrong?}

On the basis of some preliminary reports~\cite{edwin,icarusNOOPERA} which
started to circulate some six months after the original OPERA announcement
(when the relevant manuscript by the OPERA collaboration was still going
through the refereeing process)
it seems we must assume an unnoticed systematic error was responsible for the OPERA
results. Interestingly it was not one of
the systematic errors the community had been suspicious about:
in the long chain of contributions to the travel-time determination
that needed to be kept under control
with 10-nanosecond accuracy it simply happened
that a couple of steps (perhaps most notably~\cite{edwin}  a connection by
fiber-optic cable)
 were not performing as they should have.

This ending\footnote{Just as I was getting ready to publicize these notes, new results were 
reported~\cite{noOPERAjune8}
by OPERA and other neutrino experiments which fully settle the issue: we can now confidently assess the
present situation as one with no evidence of superluminal neutrinos.
Of course, it is still possible that when we will manage to reach even better sensitivities, with future
upgrades of these experiments, a true anomaly be found. Neutrinos may well be superluminal, but not
at $\sim 20 GeV$, at least not within a sensitivity of a few parts in $10^5$.}
to the OPERA story
came as no surprise to anyone. One should not imagine that the 200 papers on OPERA
imply physicists were expecting it to hold up. To give you a sense of the atmosphere
I could say that we were estimating the odds against OPERA were ten thousand to one.
But for many physicists one in ten thousand of contributing to a major discovery
is far better than our usual 100\% chance of doing something technically sound and elegant,
but of marginal significance.

It must be of interest to philosophy of science to appreciate why
we were estimating the odds against OPERA would be, say, ten thousand to one.
Part of this is easy: it just reflects the fact that it happens frequently in science
that a particularly striking experimental
result is suddenly reported, and then it is established that an
unknown bias or source of uncertainty affects those data.
We learn on the field, by experience, that
out of, say, a thousand instances of truly shocking experimental results
only one ends up holding up after further scrutiny.
So this brings us, say, to one thousand to one, a huge amount of {\underline{confidence against}}
OPERA. But my rough estimate is ten thousand to one, so this huge amount
of experience-based confidence against OPERA was actually only a tangible but
small contribution to our overall confidence against OPERA.\\
Where did the rest come from?\\
What was so suspicious about the OPERA result?\\
Since it seems the result was after all wrong, have we learned
 that our suspicions against the OPERA result were correct?\\
 Should we then have even more confidence
  in our suspicions
 in future similar situations?

Evidently these questions concern a form of knowledge that plays a crucial
role in how fundamental physics works. And yet it is not contained
in any single experimental
or mathematical result.
One of the main reasons
for our skepticism specifically toward the OPERA result, was the perceived strength of
indirect bounds on superluminality of OPERA neutrinos.
Indeed,
the significance (or lack thereof) of indirect bounds in fundamental physics is one
of the main issues I want to contemplate in these notes.

As background to this issue of indirect bounds on superluminality of OPERA-type
neutrinos one should take notice of the fact that,
perhaps surprisingly for researchers working outside of fundamental physics,
there had been already
a significant amount of theory work on possible superluminal speeds well in advance (some as
early as a couple of decades ago) of the OPERA
findings. This had been mainly motivated by research on the quantum-gravity problem,
particularly originating from:\\
(i) quantum-gravity-motivated scenarios
with additional (but tiny, and nearly invisible) spatial dimensions, and\\
(ii) quantum-gravity-motivated scenarios hosting a new principle such that the Planck
length, the tiny length scale ($\sim 10^{-35}m$) which we expect
to characterize the quantum-gravity realm~\cite{carlipREVIEW,gacQM100},
sets the minimum allowed value of wavelength
of a particle.\\
Since Einstein's relativity strictly predicts that there cannot be a minimum allowed
value of wavelength, some  studies had led to speculations about departures
from Einstein relativity at the Planck-length. And with extra spatial dimensions it had been noticed
that particles having access to the extra dimensions (not all types of particles would have access to the extra
dimensions) could be perceived
as traveling superluminally from the perspective of observers confined to our good old
3 spatial dimensions.

So we had pre-existing models for superluminal particles, and actually it was through
known results about these models that the superluminality of neutrinos tentatively
reported by OPERA appeared to be particularly hard to believe.
This originates mainly from the fact that the standard way to describe superluminal particles
in the relevant literature relies on
a formula
giving the dependence on momentum of the physical velocity
of a neutrino (or other particle) of mass $m_\nu$:
\begin{equation}
\mathsf{v} = \frac{c p}{\sqrt{p^2 + c^2 m_\nu^2}} + \Delta (p)~,
\label{vdeformed}
\end{equation}
where the first term is the standard special-relativistic formula
and $\Delta (p)$ is the desired new-physics correction term.
If for $p \sim 20 GeV$ one imposes $\Delta (p) \sim 2 \cdot 10^{-5} \, c$
then the OPERA claim is easily reproduced.
As shown in my Fig.~1 an awkwardly fine-tuned momentum-dependent correction $\Delta (p)$
would then be needed for compliance with bounds obtained at other ranges of momenta,
and a serious challenge is present even
confining the analysis at the same range of energies of the OPERA neutrinos.
This is due to the fact that a $\Delta (p)$ in the velocity formula must also be taken into account
in the analysis of measurements that are not velocity measurements, since it plays an indirect
role in the way we compute in particular the cross section for particle-physics processes.
And when this is done one finds~\cite{cohenglashow,bietal,cowsiketal} large
anomalies for particle-physics
processes, large enough to contradict known experimental facts: if the new effects
are strong enough (if $\Delta (p)$ is large enough at OPERA energies)
to produce the speed anomalies of a few parts in $10^5$ tentatively
reported by OPERA, then the same violations
also affect some well-studied particle-physics
processes at a level that contradicts our established experimental results
for those processes~\cite{cohenglashow,bietal,cowsiketal}.

When this line of reasoning was brought to full fruition
(and we were only less that a month into the OPERA-anomaly
season~\cite{cohenglashow,bietal,cowsiketal})
it had a strong effect on the attitude toward OPERA of most theorists.
It reached the point that some physicists
felt\footnote{It may be of some anecdotal interest that the authors
of these papers ``refuting OPERA", in Refs.~\cite{cohenglashowV1,icarusREFUTE},
include two Nobel laureates.}
that the OPERA result could be ``refuted"~\cite{cohenglashowV1,icarusREFUTE}
on the basis of such arguments focusing on indirect implications.

\subsection{Trying nonetheless with superluminal neutrinos}
As I shall stress more forcefully later in this manuscript,
indirect bounds have some limitations. The case against OPERA built in
Refs.~\cite{cohenglashow,bietal,cowsiketal} was ``convincing enough" according to some physicists,
but was not conclusive. And this became of interest to some of the physicists contemplating
the OPERA anomaly. After all we had known from the very beginning (at least ``thousand to one")
that the most likely hypothesis would be that the OPERA anomaly was an experimental artifact.
So from the very beginning we were just looking for a second most-likely hypothesis.
Such a second-most-likely hypothesis would have been a tool for investigating
the OPERA anomaly, if it happened to stand up long enough. For example it could have guided
our searches of some (possibly indirect) confirmation of the OPERA anomaly. Such searches
could have progressed alongside the main effort scrutinizing the
result by repeating the same measurement. From this perspective results such as those
in Refs.~\cite{cohenglashow,bietal,cowsiketal} did not change the situation very much:
they simply gave us additional reasons to think that the OPERA result could only be correct
if a particularly subtle new class of processes was being discovered (conclusion of the
same type as the one I here summarized in Fig.~1). And our experience tells us that
this does not happen often, but of course it can (and occasionally does) happen.

The most visible limitations of the case against OPERA built in
Refs.~\cite{cohenglashow,bietal,cowsiketal} originate from the fact that
the model used for deriving the argument requires departures from special relativity
involving a preferred frame. Those models stand to special relativity just like the models
of Lorentz and others stood to Galilean relativity: they required a preferred ``ether frame".
One way in which this could be exploited for going around the point made
by Refs.~\cite{cohenglashow,bietal,cowsiketal}
was used for example in studies such
as the ones in Refs.~\cite{noglashowLIVma,noglashowLIVgardner,noglashowLIVbezurkov}:
once we allow for a preferred frame to arise than we have room within the theory
for allowing, together with the speed law (\ref{vdeformed}), also other anomalous features.
Combining (\ref{vdeformed}) with other features allowed by a preferred frame setup
one could
have~\cite{noglashowLIVma,noglashowLIVgardner,noglashowLIVbezurkov} superluminal particles
with bearable consequences for well-studied cross sections.

And one could also try something more ambitious: from a relativistic perspective
it is of particular interest to contemplate the possibility
that the speed-of-light limit so far confirmed experimentally might be violated
in some appropriate regimes and yet the theory be still fully relativistic,
without a preferred ``ether" frame.
For this too there were scenarios studied in the literature
well before the OPERA anomaly
(see, {\it e.g.}, Refs.~\cite{gacdsr1,leedsrPRL,dsrnature,jurekDSRnew}),
and motivated by aspects of the quantum-gravity problem.
What is here relevant is that it had been shown~\cite{gacdsr1} that
superluminal speeds are possible preserving the relativity of inertial frames
but they should necessarily be accompanied by a modification of the law of conservation
of momentum\footnote{More precisely the law of composition of momenta must be deformed
in momentum-dependent fashion
if the speed law is deformed in momentum-dependent fashion. This is a standard situation
for relativistic theories, where the overall structure must be balanced to avoid
the presence of a preferred class of inertial observers.
In particular, we are all familiar with the fact that a speed-dependent deformation
of the Galilean velocity law can be implemented in consistently relativistic
fashion (then replacing Galilean relativity with special relativity)
only at the cost of also having a velocity-dependent deformation of the
law of composition of velocities.}.

So for example in computations of lifetimes the evaluation
of the contribution from a given two-body particle decay,
when including the effects of
the
presence of a term $\Delta(p)$
in Eq.~(\ref{vdeformed}),
must necessarily also include a deformation of the momentum-conservation law,
generically of the type
$$p^{(I)}_\mu
= p^{(II)}_\mu + p^{(III)}_\mu + \Delta_{{\cal P}}(p^{(I)}_\mu, p^{(II)}_\mu , p^{(III)}_\mu)~.$$
And the relationship between $\Delta(p)$
and $\Delta_{{\cal P}}(p^{(I)}_\mu, p^{(II)}_\mu , p^{(III)}_\mu)$ imposed~\cite{gacdsr1} by the
requirement of absence of a preferred frame ensures automatically that
the prediction for the lifetime is affected only in minute way by the deformation.
This was reanalyzed from the OPERA-neutrino perspective
in Refs.~\cite{operaDSR,fransDSROPERA,yiDSROPERA,dimitriDSROPERA},
showing that the argument for ``refuting OPERA" discussed in
Refs.~\cite{cohenglashow,bietal,cowsiketal} only applies to theories
with a preferred frame.
So within departures from special relativity preserving the equivalence of inertial frames
one could at least try to accommodate the OPERA anomaly.
And yet also this scenario encountered huge difficulties
 for producing effects such as the ones needed for the OPERA anomaly.
In particular, for this no-preferred-frame option the requirement
 of matching other data on neutrino speeds
 is rather severe. The delicate balance between modification of the speed law and modification
 of the law of conservation of momentum is hard to achieve, and the few cases that
 are known to be suitable do not accommodate complex (``ad-hoc") forms of dependence
 of speed on energy-momentum of the type shown in my Fig.1.

As mentioned above superluminal particles could also be accommodated
in the recently fashionable scenarios (much work on this for the last decade or so)
 with ``large" extra spatial dimensions (as ``large" as $10^{-20}m$),
but that too was ultimately found to be not so well suited for
 accommodating the OPERA anomaly.
At first the most promising option of this sort appeared~\cite{operaLED,operaLEDexperts,operaSTERILEnew}
 to be one where only ``sterile neutrinos"\footnote{Let me briefly summarize, for the benefit of those
 readers who are unfamiliar
 with those notions, the most significant properties of ``active neutrinos"  and of ``sterile neutrinos".
 The only neutrinos we have solid experimental evidence for are active neutrinos. Active neutrinos are indeed
 part of the presently reigning Standard Model of particle physics. These neutrinos are ``active" because,
 in addition to the tenuous gravitational interactions, they take part in the weak interactions. They are not as ``active" as electrons, which in addition take part in electromagnetic interactions, and they are much less ``active"
 than quarks, which take part in all interactions including the strong interactions, but as far as it goes for
 neutrinos they are at least active in the weak-interaction sense. The other possibility for neutrinos are ``sterile
 neutrinos" for which we only have at present indirect and inconclusive evidence. A sterile neutrino does not even
 take part in the weak interactions (so it is only subjected to the gravitational interactions, which are by far the
 weakest of all interactions).} could have access to the extra dimensions.
The muon neutrinos produced and detected at OPERA are evidently
standard ``active neutrinos" which, in addition to gravitational interactions,
also experience the weak force, and it is through weak-force interactions that they
are produced and detected. These known particles are weakly interacting at low energies
because they do not carry electromagnetic or strong-interaction charges, and at low
energies both gravity and the weak force are very weak. There is no robust
evidence of the existence of sterile neutrinos, but it is a conceivable
and much-studied (even pre-OPERA) variant of neutrinos, which only interacts gravitationally.
For them also the weak interactions would be mute.

For sterile neutrinos there were theoretical results, well in advance of
the OPERA announcement, on mechanisms
that could endow them with effectively superluminal properties, essentially
by using the geometry of the extra dimensions as an opportunity for shortcuts,
available only to sterile neutrinos.
It was immediately clear that for this scenario the
main challenge would reside in the fact that what OPERA studies are active neutrinos.
The only opportunity for involving a superluminal sterile neutrino
in the OPERA story comes from a second layer of speculation: quantum mechanics
allows certain types of particles to ``oscillate" into one another, so that
one might envisage an active neutrino
produced at CERN and being detected as an active neutrino at LNGS but traveling
from CERN to LNGS with properties that in part reflect those of a sterile neutrino,
including the conjectured superluminality. But this is easier said than done:
working out the mathematics it was found
that this scenario could produce an effect comparable to
the one reported by OPERA only at the cost of introducing rather intense active/sterile
neutrino ``oscillations", too intense for complying with the constraints on such oscillations
from other experiments~\cite{operaLEDexperts}.

In summary also the attempts of accommodating superluminal OPERA neutrinos
in frameworks not subject to the concerns reported in Ref.~\cite{cohenglashow,bietal,cowsiketal}
encountered very severe difficulties.
And this brings me back to the issue of whether and how theoretical work is
relevant in assessing an experimental result.
It is striking that on the basis
of the evidence I here summarized theorists were concluding
(even before any clear indication of an overlooked systematic error)
that describing
the OPERA claim might be hopeless.
Especially since  some sizable overlooked systematic errors
were indeed eventually found~\cite{edwin,icarusNOOPERA},
this OPERA story leaves us with the challenging question of whether
such a supremacy of theory over experiment might be possible: can an experimental result of a type not attempted previously
(not exactly the same measurement) be refuted by theorists on the basis of the fact that they
cannot find a theory that matches the relevant experimental result with other known
experimental facts?

\subsection{Still a third road: non-superluminal OPERA neutrinos}
My philosophy-of-science data reporting on OPERA would be seriously
incomplete if I did not add some remarks on studies which considered a third possibility:
non-superluminal OPERA neutrinos.
The obviously most-likely possibility was that the result by OPERA would
not be confirmed (and indeed it turned out to be right).
The second road to the OPERA was the one described in the previous subsection:
taking OPERA at face value and looking for ways to describe theoretically some
superluminal neutrinos of that sort.
But it is indeed very interesting from the viewpoint of philosophy of science
that even taking the OPERA result at face value we could not automatically
conclude that superluminal neutrinos had been discovered, at least not using exclusively
the data reported by OPERA.

One of the main reasons for this resides in the nature of the OPERA measurement:
a very intense beam of neutrinos was prepared at CERN and of the large number of
neutrinos crossing the OPERA detector at Gran Sasso only a very small fraction
could be observed (due to the weakness of the interactions of $\sim GeV$ neutrinos).
We know from other areas of physics that in certain experimental setups of this sort,
when a small sample of the particles prepared is observed, nominally (but not
substantially, see later) superluminal speeds may be measured.
Arguments based on these points and/or some related points were discussed
in a few OPERA papers (see, {\it e.g.}, Refs.~\cite{brustein,ahlu,morris}).
I am here going to focus on my personally favorite variant of this class of
possibilities, whose discussion involves topics of interest for
philosophy of science not only in relation to the OPERA story but also beyond it.

I will mainly focus in this subsection on the fact that,
even assuming Einstein relativity
to still hold, detecting some particles with a superluminal travel time from
their point of emission is not automatically excluded.
Indeed Einstein relativity predicts that no ``signal" (no information) can travel superluminally
and this translates into a bound on travel times for single particles only in the classical
limit. Within quantum mechanics
the relationship between signal and particle is complicated by wave-particle duality
and in general the notion of travel time becomes very subtle~\cite{stein2007,landauerNATURE}.

I find that these subtleties are best explained
by quickly summarizing the issues, the experimental facts and the emerging
theoretical understanding of the most famous ``travel-time issue", the one concerning
the tunneling time.
In the classical limit a  particle encountering a potential barrier higher than its kinetic
energy
simply cannot manage its way to the other side of the barrier.
Quantum-mechanical effects provide such a particle with
a non-zero (but small) probability of ending up on the other
 side of the barrier.
 This is well known and well understood. But how much time does it take a particle to quantum-tunnel
 through a barrier? This is a tough question, whose importance has been clear since the
 early days of quantum mechanics~\cite{earlytunnel1,earlytunnel2,bohmbook}, and remains
 the main focus of a very active area of both theory and experimental
 research~\cite{stein2007,physrepREVIEW,recamiREVIEW,steinbergPRL,abruptNATURE}.

One can easily see that the answer is subtle. For example, for
speeds much smaller than the speed of light
we express the speed $\mathsf{v}$ of a particle in terms of its kinetic energy $K$ and its mass $m$,
and the kinetic energy is in turn obtained subtracting to the total ``nonrelativistic energy" ${\cal E}$
the potential energy $U$:
$$\mathsf{v} = \sqrt{\frac{2K}{m}} = \sqrt{\frac{2({\cal E} - U)}{m}}~.$$
Since in quantum tunneling ${\cal E} -U <0$ this recipe for the speed (and therefore the corresponding
derivation of the travel time) becomes meaningless. This prepares us for surprises, but still does not
tell us how much time it takes a particle to quantum-tunnel
 through a barrier.

 And, as summarized in Fig.2, what we find experimentally in trying to determine the
 travel time of particles through such a quantum tunnel is rather subtle.
We have robust experimental evidence of the fact that, under appropriate conditions,
a particle prepared at time $t_i$ in a quantum state centered at (with peak of the probability
distribution located at) a certain position to one side of a barrier
is then found~\cite{stein2007,physrepREVIEW,recamiREVIEW,steinbergPRL}
 on the other side of the barrier at time $t_f$ with distribution peaked at a distance $L$
 from the initial position, with $L$ bigger
 than $c(t_f-t_i)$.
It is also by now well established that this apparently ``superluminal" behavior which we find in
 experimental data is not in conflict with Einstein's relativity, and indeed it follows
 exactly the predictions of theoretical models of quantum tunneling faithfully based
 on Einstein's relativity~\cite{stein2007,physrepREVIEW,recamiREVIEW,steinbergPRL,abruptNATURE}.

\begin{figure}[htbp!]
\includegraphics[width=0.9 \columnwidth]{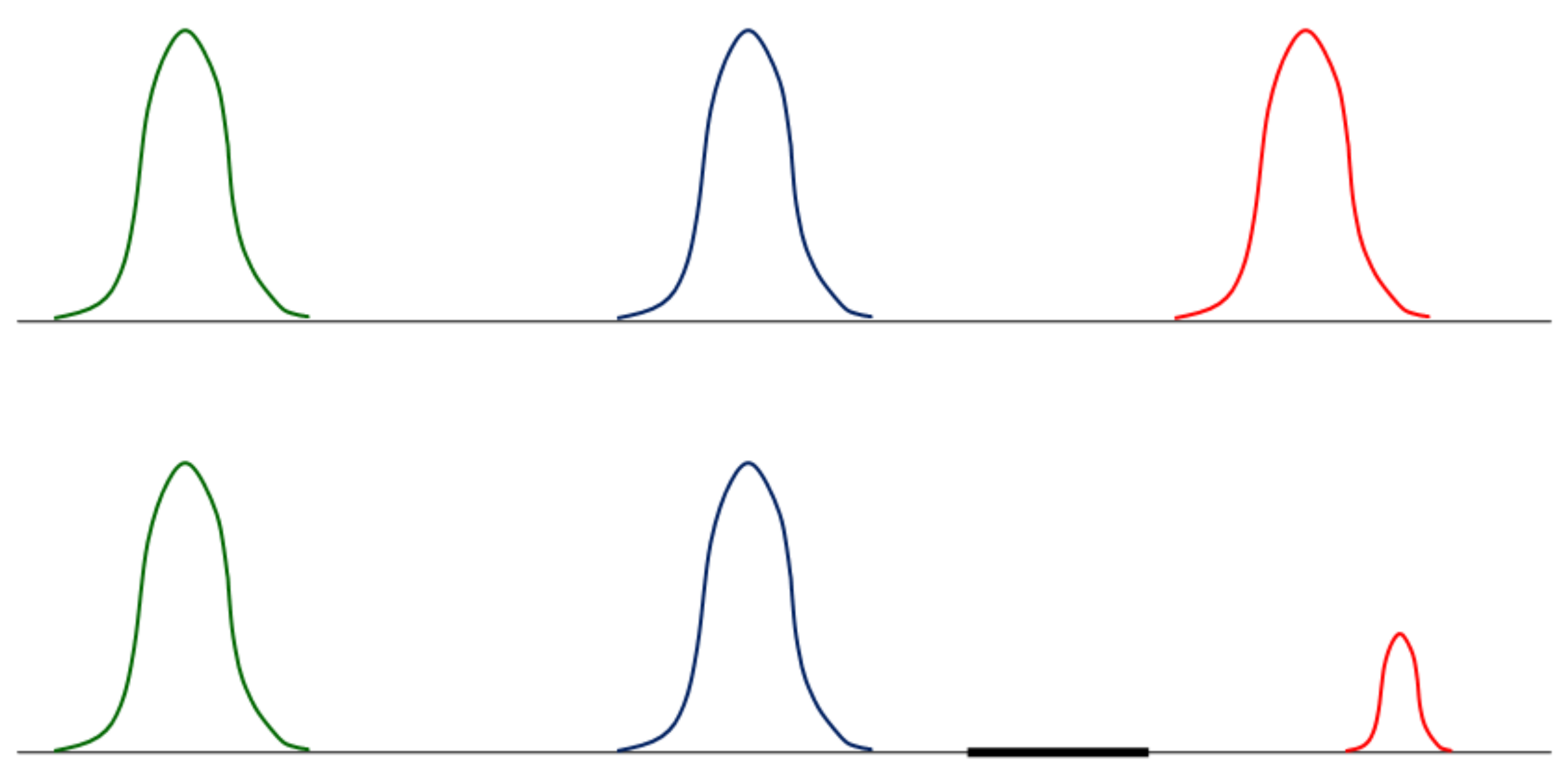}
\caption{A ``cartoonist impression" of an aspect of the tunneling-time issue.
The figure shows a time sequence (greentime, bluetime, redtime) of descriptions
of a probability distribution for the position of a particle along a given
spatial direction (the horizontal direction). Two time sequences are
shown: the top panel is for propagation without a barrier while the
 bottom panel accounts for a barrier located (but not shown in figure)
where the horizontal line is thicker.
 At redtime of the bottom panel
I am  only showing the small
transmitted peak, since showing the large
reflected peak would affect the visibility.
An anomaly appears to be present if one focuses
on the time interval found timing the peak ({{of the entire distribution}})
before reaching the barrier and timing the peak
({{of the small transmitted distribution}})
after tunneling through the barrier. Naively this would suggest
that the  particles
that do tunnel through take a time for traversing the tunnel which is shorter than the
time needed by the same particles to traverse a distance in vacuum equal to the
length of the barrier.}
\end{figure}

The key point for this emerging understanding is that such measurement setups,
determining the most likely spacetime event of production of a particle and the most likely
spacetime event of detection of a particle, establish a time interval associated to the
peaks of two different distributions. The expression ``travel time of the peak"
is only legitimate to the extent that at first
there is a large peaked distribution approaching the barrier from one side,
and then a much smaller (transmitted)
peaked distribution
is measured on the other side of the barrier. But, contrary to our instinctive
 classical-limit-based intuition, as a result of quantum-mechanical effects (such as interference)
 the peak observed after the barrier is not some simple
  fraction of the peak that was approaching the barrier. Travel times of distribution peaks,
are not travel times
of any signal, and therefore ``superluminal" experimental results for such travel times are
not in conflict with Einstein's relativity.
For smooth, frequency-band limited, distributions
the precursor tail of the distribution
allows to infer by analytic continuation~\cite{stein2007,steinbergPRL,abruptNATURE}
the structure of the peak.
Under such conditions, even for
free propagation, by the time the peak reaches a detector it carries no ``new information"~\cite{stein2007}
 with respect to the information already contained in the precursor tail.
An example of ``new information" is present in
modified distributions containing ``abrupt" signals~\cite{stein2007,abruptNATURE},
and indeed it is found that when these new-information features are sufficiently
sharp they never propagate superluminally~\cite{abruptNATURE}.

So, at least at the qualitative level, there is no such dramatic
difference between the established experimental results on tunneling times
reported, {\it e.g.}, in Ref.~\cite{steinbergPRL},
and the travel-time results that were being reported by OPERA.
But computing the signal travel time
properly for OPERA neutrinos, travelling through rock and interacting with nucleons via
 weak interactions, is much harder than for idealized quantum-tunneling exercises.
 Moreover, from this information-theory perspective OPERA's claim appeared to be quantitatively
 unmanageable:
 one can estimate~\cite{morris} that the probability distribution of an OPERA neutrino is spread out
 over roughly $10^{-3}$ nanoseconds, so anomalies for travel times of distribution peaks
  of $\sim 60$ nanoseconds (the ``travel-time anomaly reported by OPERA)
  would require an overall shift of the distribution peak by some $10^5$
  peak widths.

Of course, also from this travel-time perspective the emerging revision~\cite{edwin}
of the OPERA claim puts matters to rest. But the shared fascination for the OPERA anomaly had
the positive effect of creating some shared knowledge among particle-physics experts
of neutrino phenomena and relativists interested in travel time anomalies of the sort of
the tunneling-time issue. Moreover, pondering on the OPERA anomaly
one stumbles upon a quantitative argument suggesting that perhaps
we are not far from the point where in neutrino studies of this sort, quite independently
of the OPERA story, we might encounter some manifestation of the peculiarities
of the travel-time concept in quantum mechanics.
This estimate I can offer is based on the
well-established understanding~\cite{stein2007,physrepREVIEW,recamiREVIEW}
 that
the peculiarities of the tunneling-time problem are directly linked to the ``evanescence"
produced by the barrier: inside the barrier, in the classically forbidden region, the probability of
finding a particle decreases exponentially with the depth of the barrier.
There is some evanescence also in the description of the propagation
 of neutrinos in rock: the probability of finding a neutrino after a certain distance travelled
 in rock decrease slowly but with an exponential law, because of the small probability
 of a weak-interaction process between the neutrino and the nucleons composing the rock.
 For previous neutrino travel-time experiments~\cite{preminos1979} the evanescence of
 neutrinos in rock was neglected, and it proved to be a robust choice. And evidently,
 on the basis of Refs.~\cite{edwin,icarusNOOPERA} also for OPERA neutrinos this evanescence is
 negligible. To see that however we might not be far from the point where evanescence should
 be taken into account let me actually compute the
 fraction of a pulse of $\sim 20GeV$ neutrinos that when sent
from CERN to LNGS would be absorbed in rock. This is obtained~\cite{liparilusignoli,pdg}
multiplying
the neutrino-nucleon cross section $\sigma_{(20GeV)} \simeq 1.4 \cdot 10^{-37}cm^2$,
for the the density $\rho N_A$ of nucleons in rock,
and the distance $L$ between CERN and LNGS:
\begin{eqnarray}
\!\!\!
\!\!\!\!\!\!\!
\!\!\!\!\!\!\!\!\!\!\!\! L ~(\rho N_A) ~\sigma_{(20GeV)}  \simeq
   (7 \! \cdot \! 10^7cm) \frac{2 \! \cdot \! 10^{24}}{cm^3}
  (1.4 \!  \cdot \!  10^{-37}cm^2)  \simeq 2 \cdot 10^{-5}
 \nonumber
\end{eqnarray}
So it happens to be the case that the OPERA setup is just bringing us to the
level of accuracy of travel-time determinations where absorption-induced evanescence could
be significant:\\
 OPERA measures the CERN-LNGS travel time with accuracy of a few parts
in $10^5$ and for OPERA neutrinos  evanescence is an effect
worth $\sim 2 \cdot 10^{-5}$.

If not for the cost of such experiments it would certainly be interesting from this
perspective to have some sort of upgraded OPERA capable of achieving an even better ratio
between neutrino evanescence and travel-time accuracy. If anything of any interest
came out in this direction there I would have my centauro-event like OPERA story
(see the opening remarks of this section).

\section{Very preliminary data analysis}
\noindent
I deliberately do not attempt a detailed analysis of the philosophy-of-science data
I have here reported.
This choice not only reflects the fact that, as an amateur philosopher,
this would be beyond my strengths, but also reflects my general views on data reporting
(even for the very different context of physics data): I feel experimentalists
give their maximal contribution when they report bare data, without contamination
of theory speculations (and I am here acting as an experimentalist for the phenomenology
of philosophy of science).

Still in this section I do offer some comments on the data I reported, but my readers
should notice that here the real objective is not developing theory but rather
to use my OPERA data as basis for giving a less vague idea of what could constitute
the objectives of phenomenology of philosophy of science.

\subsection{Falsifying theories?}
It seems among modern philosophers the idea that theories can be falsified
is out of fashion, but remains a key reference point in the background of the philosophy-of-science debate.
The OPERA story provides a perfectly suitable context for articulating viewpoints
related to this issue on the basis
of contemporary developments rather than digging (with all the ``fuzziness" that implies)
in the history of physics\footnote{Moreover, for some issues of interest for philosophy of science,
such as understanding the ingredients of a ``scientific revolution", I feel that focusing exclusively
on paradigm shifts that did occur is limiting. There is as much to be learned on these issues from
cases where a ``scientific revolution" was tried but failed (as in this OPERA story)
as in cases where actually the proposed paradigm shift was ultimately adopted. I find that historians
of physics introduce a bias in this respect by giving much more accurate accounts of
the paradigm shifts that did prevail with respect to the cases of failed attempts of paradigm shift.}.

Actually I feel that the most interesting aspects of the OPERA story for what concerns the falsification
issue reside in the fact that the OPERA experiment was testing a simple ``law", the
relativistic bound $v\leq c$ describing the speed of light as the maximum achievable speed.
This affords us some freedom from the cumbersome aspects of ``falsifying a theory"
or ``replacing a theory paradigm". And in turn this is comforting to me:
I surely
do not read enough philosophy (and I might have been reading the wrong sources),
but it happens to be the case that most
of the philosophers I did read use a very naive notion of ``falsifying a theory"
for which I find no place in science.
Evidently the core activity of science
is to summarize large collections of measurement results into some useful/predictive
simple ``laws" and some useful/predictive manageable algorithms\footnote{While, as emphasized above,
with these notes I am not intending to do any ``theory of knowledge" or ``theory of science",
let me take the brief liberty
to stress that what I gave as description of the core activity of science does not amount to ascribing
to the philosophy of instrumentalism. As a humble theoretical physicist
(and a humble experimentalist of philosophy of science)
the concept of ``true picture of the world" is unintelligible to me, but I am well prepared to acknowledge
that others may be capable of appreciating this concept and dwell about it.
Still my description of the core activity of science as the one of
summarizing large collections of measurement results into some useful/predictive
simple ``laws" and some useful/predictive manageable algorithms is evidently correct, factually correct.
Dwelling on whether those typical products of science contain (or bring us any closer to)
the ``true picture of the world" is a task better reserved to philosophers (theorists of knowledge).
So evidently
I am not advocating the philosophy of instrumentalism, but I am advocating a probably novel approach
in which the answers to questions such as ``how and why science works?" are sought in two stages: a first
stage where concerns  for the "true picture of
the world" are postponed and
one adopts an instrumentalist perspective (not a full philosophy, but just a perspective for the first
level of analysis), and a second stage where issues pertaining to what is the "true picture of
the world" come into discussion.}.
The simple law $v\leq c$ is a prototypical example of what a very powerful law can do for us.
There is no higher achievement in science than establishing for such a powerful law
the limitations to its
applicability (which is the situation usually described, with horrible choice of terminology,
as ``falsifying" the law).

But even shifting our focus from the overwhelming goals of ``falsifying theories"
and/or ``replacing paradigms" to the more limited arena of ``falsifying laws" of
the type $v\leq c$ the challenges for philosophy of science (and, in my view, for the
phenomenology of philosophy of science) are not completely trivial.\\
Can we really falsify $v\leq c$? in which sense? on the background of some
reigning theory and/or reigning paradigm? or in some more theory-independent sense?\\
Relevant for exploring conceptually these issues is the exercise of
imagining the hypothetical situation in which the OPERA result actually
stood up, and repetitions of the OPERA measurement confirmed it.
Also relevant is the fact that we do have the technological ability for
performing an OPERA-type measurement in (good) vacuum  rather than for neutrinos traveling
through rock
(it would require a huge money investment but the technology is there for doing it).
What if the OPERA result was confirmed also in vacuum? And testing it over a broad range
of distances we kept finding speeds of $\sim 20 GeV$ neutrino that were superluminal
by $\sim 2 \cdot 10^{-5}$?

What would it take for us to falsify the claim that there are no superluminal speeds?

There are many (far too many) pages of philosophy of science devoted to whether
Newtonian gravity had been falsified at the time when Einstein gravity was adopted.
Embracing phenomenology of philosophy of science also means replacing some of those
pages with pages devoted, for example, to the OPERA anomaly attempting to establish
what it would take for OPERA-type measurements to falsify the speed-of-light limit.

\subsection{Refuting experimental results?}
As I stressed repeatedly, the OPERA story is an example of
attempt by theorists to {\underline{refute an experimental result}}.\\
Can that be done?\\
The philosophy-of-science data I have reported suggest that it can,
but I believe they are inconclusive.\\
Actually, as an old-fashioned physicist I stick to the idea that
theorists can never refute experiments. An experimental result can only be refuted by other
experimental results
 (and only to a certain extent).
An example of what I mean by other experiments refuting a certain given experimental result
is provided by the relatively recent story concerning cold fusion: a huge effort was directed
toward reproducing the initial claim by Fleischmann and Pons in other experiments, and the evidence
gathered by those studies allowed to establish that some source of systematic error had to have
been overlooked in the Fleischmann-Pons study. In such cases we might even never know which specific
systematic error had been overlooked and yet establish, on the basis of the
consistent results of other identical (or nearly
identical) experiments, that there had to have been some overlooked systematic error in the
original Fleischmann-Pons study. Fact is the Fleischmann-Pons study did not lead to the introduction
of some useful/predictive
simple ``law" or some useful/predictive manageable algorithms. Humanity has no cold fusion (as of 2012).

So experimental results can be refuted\footnote{Let me be perhaps overzealous and
stress the difference between the notion of ``refuting an experimental result" 
and the notion of ``reinterpreting
an experiment". I appear to find that in philosophy publications the notion of
reinterpreting
an experiment is of very strong interest, often even more than the notion of refuting an experimental
result.
 This will always be the case when philosophy dwells about the
the ``true picture of the world", since reinterpretation is a crucial issue for such debates.
An example of reinterpretation is provided by comparing the {\underline{description of}} experimental
results on the gain or loss of weight by materials being burned that was fashionable at the time
of the Phlogiston Theory and the {\underline{description of}} those same experimental
results that became fashionable after the discovery of oxygen. The discovery of oxygen in no way affects
the robustness of previous experimental results on the gain or loss of weight by materials being burned,
and in particular the discovery of oxygen in no way led to the discovery of unnoticed sources of systematic
error in those experimental results. The same experimental results apply equally well to the theories informed
by the discovery of oxygen but the language used in describing those results has changed.
Experiments done nowdays on the gain or loss of weight by materials being burned still find
results for those changes of weight that are fully consistent with the ones from 3 centuries ago.
In contrast no issue of
reinterpretation is involved in the Fleischmann-Pons cold-fusion story. The experimental result
announced by Fleischmann and Pons was refuted by performing other experiments identical to (or nearly
identical to) the Fleischmann-Pons type. If nowadays we follow the procedures described 3 centuries ago
for inducing changes of weight of materials by burning we indeed obtain changes of weight of materials by burning.
If nowadays we follow the procedures described some 20 years ago
by Fleischmann and Pons for obtaining excess heat we do not obtain excess heat. Next century probably what
Fleischmann and Pons were trying to produce will not be interpreted as excess heat but as something else,
and yet
if at some point in the next century someone follows the
Fleischmann-Pons procedure for producing excess heat (then reinterpreted as something else)
that someone will still not obtain any excess heat (unless the relevant laws change
wildly over time...).} by other experimental results.
The view that this is the only way we can refute an experimental result
is not a consensus view among physicists, as shown by aspects of the OPERA story which I summarized above.\\
Can philosophy of science help?
Can theorists, at least in some very special situations,
refute an experimental result, even before any information is available from experiments
 attempting to reproduce that experimental result?

\subsection{Blinding and the theory of knowledge}
One point that should not be missed in the phenomenology-of-science data I reported
is that the efforts motivated by the (apparent)
OPERA anomaly produced ideas worth exploring and made us learn facts worth knowing.
Some of the techniques of data analysis invented by the OPERA collaboration
were subjected to very carefully scrutiny and they proved reliable.
These techniques are now a valuable resource for the future. This in particular applies
to the case of the techniques used by the OPERA collaboration for extracting
an estimate of single-particle travel times from a setup which required
a very subtle statistical analysis.

And also some of the theory results obtained studying the OPERA anomaly have
significance that extends well beyond the short season of the OPERA story.
For example, some of the results obtained for violations of special relativity
in the OPERA regime also apply to studies of the fate of relativity
at the Planck scale, where we expect a rather virulent interplay
between quantum mechanics and general relativity with potentially significant
implications for tests of special relativity~\cite{gacdsr1,leedsrPRL,dsrnature,jurekDSRnew}.

I find that the overall balance of the OPERA story is positive.
But some other physicists argue
that the OPERA story should teach us to better
guard against the wastefulness of (the wastefulness they perceive in)
questioning a sound law.
I fear that what might ultimately be at stake here in this useful/wasteful
difference of assessments is the fate of the blind-analysis standards
in fundamental physics.
As in most modern particle-physics experiments, OPERA's analyses were ``blinded":
the criteria used, including the estimates of systematic errors, were fine-tuned before the data were looked at. Once the data were analysed, the results were announced without much delay and with no further tweaks.
Experience tells us that strict adoption of this blind-analysis procedure
maximizes our chances for new discoveries and for sound reporting of experimental results.
And an analysis of such concerns must be relevant for
any ``theory of knowledge".

However, if indeed the preliminary OPERA result is not confirmed,
some are bound to propose that we soften the blind-analysis standards.
Experimentalists who find that their results contrast with ``known" physics might be encouraged to postpone announcement of the results, and first look determinedly, in a non-blinded way, for systematic errors that might make the contrast go away.
This is horrifying to me, since I fear it would introduce a potentially disastrous bias against important discoveries. Questioning our laws, even on the basis of preliminary experiments, is a healthy exercise. And we should have as our top priority the one of not introducing
any risk of missing a potentially
important discovery.

On the contrary the OPERA story might leave in fundamental physics not only effectively
lowered standards of blind analysis, but also an atmosphere hostile to new discoveries.
It must be already difficult for a young experimentalist to report results that appear
to contradict ``known" facts. And it will be now perhaps even more difficult.
For example, the PI of the OPERA experiment was led to resign its role.
What sort of message does that convey?

I hope the sort of message I imagine will not reach young experimentalists with
good innovative experimental ideas. But surely an opportunity was missed with
the general public: the OPERA story attracted huge attention in the general public
and as it unfolded (particularly with the news of the resignation of OPERA's PI)
the general public got the idea that this was science done wrongly,
whereas it was just a prototypical example of science at its best
(though not its luckiest).

\subsection{On-time progressiveness/degenerativeness analysis}
I am arguing that some aspects of philosophy of science are part of science,
they are useful to mankind and the relevant theories can be tested.
For example the best theories of knowledge will survive confrontation
with the type of data I have reported, and weaker alternatives
will be ``effectively falsified".

Theories of knowledge that are useful to mankind should for example guide
our choices about which research programmes should be funded by citizens who
have chosen a certain set of objectives for their society.
An example of philosophical debate that is relevant from this perspective
is the one mainly centered around Lakatos notion of progressive and degenerative
research programmes. No trace of this is found in the way funding agencies
evaluate research programmes in fundamental physics,
and probably this is causing a tangible waste of resources.

The analysis of these aspects of the
theory of knowledge cannot be confined to
endless highly-educated debates. We must test them.
And clearly we can, at least to some extent we can test them: if these
(aspects of) theories of knowledge are
supposed to be useful there must be a way for us to verify whether or not a given proposal
for such a theory is indeed useful. [Here the word ``useful" is evidently not free from challenges
of objectification and quantification, but shying away from those challenges, perhaps only because
of discomfort toward seeking partial/improvable answers, is an option we cannot afford, considering
the limitations of our resources.]

I notice that to some extent this is done, but only in the awkward sense of postdictions:
alternative theories of knowledge compete on the basis of how well they describe
what happened in the past. But if they are to be useful
they better be good at describing the future, making predictions,
as normal for any science.
And this is where ``experimental tests" would be relevant within phenomenology
of philosophy of science.
Let me take the OPERA story and the notions of progressive and degenerative research
 programmes as an example. The OPERA literature was composed of
more than 200 papers in 6 months, with theoretical-physics theories born and dead in a matter
of weeks. In order for phenomenology of philosophy of science to work one would
need a comparably sizable and comparably timely\footnote{A related
perspective can be found in Ref.~\cite{rickles}, where it is observed that the ongoing
development of research on the quantum-gravity problem deserves being of interest for philosophy of
 science right now, as the process unfolds (rather than being postponed until ``quantum gravity is found"
 and some sort of consensus on quantum gravity is reached).} effort on the philosophy-of-science side,
over those six months, providing for example
analyses of which OPERA-anomaly theories were, at a given time, in a degenerative or progressive
phase. Another example would be the analysis of studies claiming
 that the OPERA result could be refuted and/or that certain
theoretical descriptions of the OPERA anomaly could be ruled out.

\subsection{``Normal science" and the weakness of indirect bounds in fundamental physics}
``Normal science" is a notion from philosophy of science which is totally foreign to this author.
I know only good and bad science.
At least in fundamental physics one can however find applicability
for a notion of ``normal-science season" which could be used to
identify a period in which no experimental result is striking enough
to force a change of ``theory paradigm". It is true that the history of fundamental
physics could be divided accordingly into a sequence of normal-science seasons
interrupted by periods where a new theory paradigm is established.

The OPERA phenomenology-of-philosophy-of-science data I here reported clearly testify
that in the present (depressingly long) normal-science season many physicists maintain
full readiness toward possibly adopting a new paradigm, as soon as experimental results
suggest\footnote{Of course, it is as usual difficult to look for a new paradigm only equipped
with the presently-adopted paradigms.
At any given time there are plenty of experimental results that do not fit in the present paradigm,
and also for this the OPERA anomaly is a good example.
But it takes time for us to be confident that one of these experimental results must be
taken seriously. Most such anomalies follow a path very similar to the one of the OPERA anomaly,
a path taking us to establish a pretty clear underestimation of systematic errors.
And yet we must (and we do) keep questioning them all, with OPERA-anomaly style.
It seems we cannot do better than questioning them from within the known paradigms
and the perhaps too cautious modifications of such paradigms which we are able to conceive.
With supernatural powers one could analyze all emerging (and usually illusory,
but occasionally magnificent) experimental puzzles in some sort of comprehensive ``space
of paradigms", with better chances of noticing quickly that a new paradigm starts to be needed.
The slower process available to us is evidently good enough.}
this might be appropriate. Several physicists working at the present time
are actually mainly focusing their work on figuring out what are the best chances
for experimental studies to finally provide a hint toward a new paradigm, and research
on OPERA was just a tiny snapshot of what for example is a characteristic feature from
this perspective of most of research on the quantum-gravity problem.

With all the pages devoted to creative reconstructions of facts occurring
in some distant past of science, history and philosophy of science would be clearly
better served diverting a good portion of that attention to properly accounting
for what is going on now. One could easily see that the present fundamental-physics
community is driven by a sort of equilibrium between those who take responsibility
for seeking the advent of a new paradigm and those who take responsibility
for preserving the status quo. Since the idea of balance in a single human scientist
remains  a wild abstraction, we have this mechanism for balance (for keeping science on the right
track) composed of two extremes: fools ready to swear we have understood everything and
wackos\footnote{Bias disclosure: I impose upon myself as work attitude an extreme wacko mode.
And on the rare occasions when I make time to read Newton and Galileo I have a pretty weird
perspective. On the one hand, when I make myself aware of the historical context where those
theories were formulated, and I remind myself of how often, even in this 21st century, we still use those
theories (for tasks not extending beyond their regimes of applicability)
I am overwhelmed with admiration for Newton and Galileo. But on the other hand,
if I allow myself to step away from awareness of the relevant historical context, and read
Newton and Galileo as if they were writing now, I am overwhelmed with ``poor-thing sympathy"
because of their amusingly clumsy way to describe Nature.
Advocates of the notion of ``theory of everything" implicitly assume that we are producing
in this season of fundamental physics the ultimate descriptions of Nature.
I am absolutely sure that a scientist of the 25th century when looking (if ever) upon the studies
we produce in the present season of fundamental physics will at best be
overwhelmed with ``poor-thing sympathy".}
ready to swear we only got to know a tiny fraction of what is out there to be known.

I would argue that some interest from history and philosophy of science is also deserved
by the missing signatures on the OPERA papers~\cite{operaOLD,opera}: a significant
fraction (though a minority)
of the OPERA collaboration did not sign these papers. This may be worth investing
some sizable investigatory work. Uncontrolled rumors easily trackable on blogs and
other internet sources suggest that some of these missing signatures are from experimentalists
that, fully consistently with the scientific method, were (instinctively or even on the
basis of direct expertise with aspects of the apparatus) feeling the announcement would be premature.
These colleagues do not share my concerns for strict adherence to the blind-analysis standards,
but this is a legitimate (though wrong) scientific position. It appears likely however that
some of those who did not sign the papers simply did not sign them because they felt
it would be foolish to ``challenge Einstein relativity". Some philosophers may label this second
mechanism for not signing the papers as an aspect of ``normal science", in my opinion
granting too much complexity to a simple manifestation of bad science (non-science).

And I would argue that of even greater interest for philosophy of science is the role
played by ``indirect bounds" in the analyses of the OPERA anomaly.
My intuition is that a certain way to use indirect bounds is characteristic of
a ``normal-science season".
A standard example of use of indirect bounds is the hypothesis of adding sterile neutrinos
to the Standard Model of particle physics. This is a fully specified hypothesis: it amounts to
a specific change to the Standard Model, adding a particle we have no direct evidence for
to the particles that are already in the Standard Model.
Such a fully specified hypothesis can be tested not only directly, by looking for
the sterile neutrinos themselves, but also indirectly, by looking for the differences
for other particles that arise if one adds to the Standard Model the sterile neutrinos
(with respect to the case without sterile neutrinos).
Bounds on some qualities of such sterile neutrinos
(such as their masses) can then obviously be obtained without observing them directly,
but rather studying more accurately properties of the Standard-Model particles we already know.

This method of indirect bounds has a long and very reputable tradition in fundamental physics,
and to some extent in the whole of science. But it should be noticed that the proposal of sterile
neutrinos that I just described involves no paradigm change: one adds sterile neutrinos to
the existing paradigm and more precisely in a very specific manner within a known theory
(a theory fully fitting with the presently adopted paradigm).
The OPERA anomaly however had the potentialities of being a paradigm shifter: if for example
it had led to abandoning special relativity it would have taken us to some place we cannot be sure
of being able to imagine right now. A crucial question for fundamental physics is ``to what extent
we can rely on indirect bounds when we analyze an hypothesis of paradigm shift?".
I leave to trained philosophers of science to ponder this question.
My uneducated position is that in such cases indirect bounds carry little or no relevance.
But, as shown by the philosophy-of-science data I reported above, it is rather
common for most physicists of this normal-science season to think that
indirect bounds also apply to cases of investigation of paradigm shift.
This is particularly clear in Subsec.~1.3: the OPERA tentative announcement
of superluminal $\sim 20 GeV$ neutrinos was being ``refuted" on the basis of the fact that
the simplest models incorporating superluminality of such neutrinos would have affected
some particle-physics processes in ways incompatible with established experimental bounds
on those particle-physics processes.

\subsection{Regimes of physics: what did we learn from OPERA?}
I already stressed above that (at least from the ``centauro-events perspective" discussed
in the opening remarks of the previous section)
we did learn a lot from studying the apparent
OPERA anomaly.

There is however an aspect of ``what is knowledge", relevant for studies of
the OPERA neutrino-velocity measurement, that feels more real, at least to my
uneducated philosophical eyes. This has to do with the notion of ``regimes of physics".
More precisely the notion of ``new regime of physics", which I shall only be able to describe
here very preliminarily. And in spite of the preliminary nature of my characterization
readers will notice that my objective is to argue that our knowledge always grows very
significantly, independently of the outcome of the experiment,
whenever we manage to perform a new experiment that gives us access
to a ``genuinely new regime of physics". I expect that the philosophy debate on this notion
might end up teaching us that each measurement gives us access to a new regime of physics,
but what is desirable for the progress of science is to somehow quantify the amount of novelty
of the regime studied by an experiment, so that we can assign priority (also for what concerns funding)
to experiments probing more novel regimes of physics.

I will illustrate part of these notions  by stressing that the OPERA neutrino-velocity
measurement did gain us access to a genuinely new regime of physics.
Most researchers working on the OPERA anomaly were particularly puzzled by it because
this was for particles with energies of a few tens of GeV, {\it i.e.} energies
which we have produced with ease at particle accelerators for the last few decades:
there appeared to be a mismatch between the amount of surprise the
(preliminary) measurement result contained and the rather ordinary regime
of physics that was being probed. This way of reasoning reflects the structure of the
extraordinarily successful ``discovery paradigm"\footnote{I am aware of the usefulness of
the notion of ``theory paradigm" within philosophy of science. With ``discovery paradigm" I am clearly
labeling a rather different, actually complementary notion. The complementarity resides in the fact
that a theory paradigm is used for interpreting/describing experimental results, whereas a ``discovery paradigm"
is used as guidance for devising new experiments perceived as providing good chances of significant
new discoveries. The most prototypical and successful discovery paradigm of 20th-century
fundamental physics was centered
on particle accelerators: we kept seeking ways to produce (and collide) particles of higher and higher energy,
and at each new energy level reached we made experimental discoveries which helped us shape the Standard Model.
 Incidentally, it appears that this discovery paradigm is running out of steam: achieving greater particle energies
  used to be easier and used to produce more significant discoveries in earlier applications of
  this discovery paradigm, whereas for the most recent applications of this discovery paradigm gaining even
  just an order of magnitude in particle energy took
  a huge effort and produced relatively insignificant discoveries.} for fundamental physics
in the 20th century: we kept making important new discoveries by gradually
gaining access to experiments producing particles of higher and higher energy.

This paradigm was so successful that at this point some physicists cannot see any other
paradigm. But I expect that every discovery paradigm eventually runs out of steam. And several
hints suggest that the 20th century discovery paradigm is at the point of needing to be replaced.
Also from this perspective it is important to appreciate that actually one can make a meaningful case
that OPERA neutrino-velocity measurements did gain us access to an interestingly new regime of
physics.

The main point for appreciating this is to notice that, although their energies have values
in no way exceptional, the speeds of OPERA's neutrinos are truly exceptional.
Among previous accurate measurements of velocity for massive particles we have the
ones in Ref.~\cite{electron1975}
for electrons with energy of about $15 GeV$. Therefore energies comparable to the ones
of OPERA neutrinos, but speeds significantly smaller than for OPERA neutrinos: indeed according
to special relativity
$$\frac{c-v}{c} \simeq \frac{m^2}{E^2}$$
so for particles of the same energy $E$ the speed is much closer to the limiting
speed of light if the mass is smaller. And for electrons we know that the mass
is $\sim 0.5 \cdot 10^{6}eV$ while for neutrinos we have so far established that the mass
is $\lesssim 10eV$.

As mentioned above, the neutrino speed measurement that had been previously
reported in Ref.~\cite{preminos1979} was for neutrinos
of higher energy, and therefore speed even closer to the speed of light,
but had lower accuracy.
So, as far as concerns measurements of speeds of massive particles (speeds therefore
inevitably different from the speed of light) the OPERA measurement was truly unique,
record setting. It is in these cases that my concerns for the use of indirect bounds
apply most strongly: how can we use indirect bounds on a regime never accessed before
and for testing the hypothesis that this regime requires a new theory paradigm? I think
one simply cannot apply indirect bounds to such hypotheses, and I am confident
that the phenomenology of philosophy of science will eventually find results
in agreement with this expectation.

And let me also stress that the regime probed by OPERA is also a regime for which
in some sense, at least adopting a strict exploratory attitude, we are prepared for
surprises. In order to appreciate this one should be familiar with the ``mystery" surrounding
the possible role of the ``Planck scale" in physics~\cite{carlipREVIEW,gacQM100}.
We expect this scale to be characteristic of the onset of some new theory paradigm.
By combining known values of the speed-of-light scale, Planck constant and Newton constant
we estimate this scale to be
$$E_P = \sqrt{\frac{\hbar c^5}{G_N}} \simeq 1.2 \cdot 10^{28}eV$$
Assuming (rather legitimately) that the mass of OPERA neutrinos is roughly of the order
of one electronvolt we have that for those neutrinos
\begin{equation}
\frac{E}{m} > \frac{E_P}{E}
\label{newregime}
\end{equation}
suggesting that the energy of OPERA neutrinos is closer to the Planck scale than it is
close to its minimum value, the rest energy $m$. For an electron condition (\ref{newregime})
requires energies of $\sim 10^{17}eV$, and we never saw an electron like that.
For a proton condition (\ref{newregime})
requires energies of $\sim 10^{19}eV$, and we infer that some of the ultra-high-energy cosmic
rays we observe are protons with energies between $10^{19}eV$ and $10^{21}eV$, but for
those ultra-high-energy cosmic rays we have absolutely no speed determination in our data,
so as far as we know their speeds could have all sorts of weird properties.

I went through this exercise of attempting to characterize the OPERA neutrino-speed measurement
as one that gained us access to a new regime also as a way to illustrate the type of reasoning
a physicist can follow in trying to assess the proposal (or the results of) an experiment.
The line of reasoning I adopted shows how these assessments can be very significant
but also suggests that they might lack objectivity:
do the crude estimates I showed above really provide an objective characterization
of the OPERA measurement as a new-regime measurement?
Or could I build a case of similar nature for any experiment if I allowed myself
to use logics so loosely?
What constitutes then a truly new regime of physics? Can we have model-independent characterizations
of what constitutes a new regime of physics? Or is it the case that
our arguments of what could be an interesting new regime of physics are so dependent on
our current models/paradigms that they loose meaningfulness in the search of new theory paradigms?

\section{A little more on phenomenology of philosophy of science}
In closing, let me look a bit beyond the OPERA anomaly. After all, evidently, I used
the OPERA anomaly as pretext for trying to make a point of more general applicability.

Fundamental physics of the past, particularly from the beginning of the 20th century,
has played a significant role in shaping parts of modern philosophy of science.
Perhaps most philosophers have not noticed how different fundamental physics is now.
It is not just that we of course study different questions, but rather we study them
in a way that is different. The OPERA-story pretext allowed me illustrate these changes
more simply than needed for other aspects of fundamental physics
that may require the urgent attention of philosophy, including ``quantum gravity",
``theories of everything"\footnote{Bias disclosure: I am on the far extreme
of ``against everything". Actually a good part of my motivation for proposing a ``phenomenology
of philosophy of science" is that I expect its development
can contrast the advance of the idea of ``theory
of everything" among physicists. I shall not dwell on this here, but I find
that faith in a theory of everything (or even just faith in the idea
of a theory of everything) is not only to be placed outside science,
as all matters of faith, but is actually armful to
science. Faith in ``theory of everything" reduces our ability
to be sufficiently inquisitive in our exploration of Nature.}, ``anthropic principle", and the ``multiverse".
My main point here was that philosophy of science can be useful for all this. It can be
tangibly useful, at least in as much as it can inform our choices concerning where
to invest resources if we are chasing a given set of objectives. But any human endeavor that
wants to claim to be useful is automatically subjectable to experimental test,
and therefore requires the development of a phenomenology. If a theory of knowledge is tangibly
useful, than alternative theories of knowledge should be compared on the basis
of their track record\footnote{While working on this manuscript, still uncertain whether
it would make sense to announce such uneducated philosophy speculations, I took
it as a good sign when I stumbled on the internet (searching for something else of use for
these notes) on evidence of how Galileo's first explorations of the scientific method
were looked down by contemporaries, as if his reasoning was so weak it needed
the vulgar help of experiments.} of usefulness.
And among the efforts needed of course we should include theories
(improvable/falsifiable theories) of what is and how we should
quantify this usefulness of theories of knowledge.
Moreover, just like with ordinary scientific programmes, we should
acknowledge the need to keep a certain balance in the distribution of resources:
daring to assign more resources to the theories of knowledge with a more established track record of
usefulness while not neglecting completely out-of-mainstream theories of knowledge (at a later
reassessment of the situation we may find that some theory previously placed out of the mainstream
has in the meantime established a better track record and should receive top-level funding
and encouragement).

Something else which is changing, and is a point of doubly-special relevance for
this manuscript, is the ``distance" between theorists and experimentalists.
Philosophy must have something to say about this. Right now in fundamental physics most theorists
understand nothing about experiments and most experimentalists understand nothing about theory.
Those who really try very hard only manage very little: some theorists manage to get
vague understanding of the workings of fundamental-physics experiments and some experimentalists
have a partial understanding of the formalisms used in fundamental-physics theories.
This situation is in part due to the nature of the task involved, but it is also encouraged by
practices in the community (again a typical example being criteria for funding) that do not
sufficiently support the efforts that some of us direct toward an interdisciplinary theory-experiment
expertise.

For the phenomenology of philosophy of science which I here ventured to imagine
there would be from the onset such a huge separation between experimentalists and theorists.
The typical experimentalist of philosophy of science would
be someone like myself, a fundamental-physics theorist with little or no understanding
of the philosophy involved, and already now it would be very hard for
a philosopher to gain a better than superficial knowledge of the theories
and experiments at the frontier of fundamental physics. Perhaps it cannot be done.
But I feel we must at least try.
Perhaps we need the guidance of some philosophy of the phenomenology of philosophy of science.

\section*{Acknowledgements}
Before publicizing these notes more widely I sent a version of them to
Dean Rickles and
John Stachel. I am very grateful to Dean and John for the very valuable  feed-back they gave me.
Some aspects of this less unpolished version which I am publicizing reflect very directly
those exchanges of ideas with Dean and John.
For example, some of the comments on  ``what is knowledge?" were added
in last-minute editing:
discussions with John made me more aware of how big  the gap was (and still is)
between my appreciation
of the subtlety of the concept of ``knowledge" and the complexity of the philosophy
debate on what and where knowledge is. Another example are my discussions with Dean about
the issue of refuting experimental results (which, attempting to be provocative, in an earlier version
of these notes I occasionally labeled as ``falsifying experiments"). And the idea of mentioning the
example of measurements concerning the changes of weight of materials upon burning (footnote 8)
came up in these discussions with Dean.


\begin{thebibliography}{50}

\bibitem{dawn} G. Amelino-Camelia, {\it Are we at
 the dawn of quantum-gravity phenomenology?},
gr-qc/9910089, {Lect.~Notes Phys.}~{541} (2000) 1.

\bibitem{isham} C.J.~Isham,
\textit{Structural issues in quantum gravity},
in \textit{Proceedings of General relativity and gravitation 1995}
(World Scientic, Singapore 1997).

\bibitem{grbgac} G. Amelino-Camelia, J. Ellis, N.E. Mavromatos, D.V. Nanopoulos and S. Sarkar,
Nature {\bf 393} (1998) 763

\bibitem{gampul}
R.~Gambini and J.~Pullin,
{Phys.~Rev.}~{D59} (1999) 124021.

\bibitem{schaefer} B.E.~Schaefer,
Phys.~Rev~Lett.~82 (1999) 4964.

\bibitem{gacgwi} G. Amelino-Camelia,
gr-qc/9808029,
{ Nature} {  398} (1999) 216.

\bibitem{mexweave} J.~Alfaro, H.A.~Morales-Tecotl and L.F.~Urrutia,
gr-qc/9909079,
Phys.~Rev.~Lett.~84 (2000) 2318.

\bibitem{jaconature} T.~Jacobson, S.~Liberati and D.~Mattingly,
astro-ph/0212190v2,
Nature {424} (2003) 1019.

\bibitem{piranNeutriNat} U.~Jacob and T.~Piran, Nature Physics 3 (2007) 87.

\bibitem{gacPRL2009}
G.~Amelino-Camelia, C.~Laemmerzahl, F.~Mercati and G.~M.~Tino,
arXiv:0911.1020, Phys.\ Rev.\ Lett.\  { 103} (2009) 171302.

\bibitem{fermiNATURE}
A.~Abdo \emph{et al.},
Nature 462 (2009) 331.

\bibitem{whataboutbob}
  G.~Amelino-Camelia, M.~Matassa, F.~Mercati and G.~Rosati,
  arXiv:1006.2126,
  Phys.~Rev.~Lett.~106 (2011) 071301.

\bibitem{stachelWHERE} J.~Stachel, ``Where is knowledge?", talk at ``11th International
Symposium on the Frontiers of Fundamental Physics; Paris, 6-9 July 2011"
(also see J.J.~Dickau, Prespacetime Journal 1 (2010) 816)

\bibitem{operaOLD}
T.~Adam {\it et al},
arXiv:1109.4897v1

\bibitem{opera}
T.~Adam {\it et al},
arXiv:1109.4897v2
	
\bibitem{gacbjorken} 	
G.~Amelino-Camelia, J.D.~Bjorken and S.E.~Larsson,
Phys.~Rev.~D56 (1997) 6942

\bibitem{centauro} 	
V.~Kopenkin and Y.~Fujimoto, Phys.~Rev.~D73 (2006) 082001


\bibitem{whataboutopera}
  G.~Amelino-Camelia, G.~Gubitosi, N.~Loret, F.~Mercati, G.~Rosati, P.~Lipari,
arXiv:1109.5172,  Int.~J.~Mod.~Phys.~{\bf D20} (2011) 2623.

\bibitem{giudice}
G.~F.~Giudice, S.~Sibiryakov and A.~Strumia,
 arXiv:1109.5682

\bibitem{operaELLIS} J.~Alexandre, J.~Ellis and N.E.~Mavromatos,
 arXiv:1109.6296

\bibitem{preminos1979}
G.~R.~Kalbfleisch, N.~Baggett, E.~C.~Fowler and J.~Alspector, Phys.\ Rev.\ Lett. { 43} (1979) 1361

\bibitem{wrong1} C.R.~Contaldi, arXiv:1109.6160

\bibitem{gpsOPERAnote}
G.~Colosimo {\it et al},
OPERA public note 132, 2nd version

\bibitem{sagnacOPERAnote} E.~Kiritsis and F.~Nitti,
OPERA public note 136

\bibitem{bigpulseISbad} G.~Henri, arXiv:1110.0239

\bibitem{edwin} E.~Cartlidge, {\it Error Undoes Faster-Than-Light
Neutrino Results}, ScienceInsider 22 February 2012
(news.sciencemag.org/scienceinsider/)

\bibitem{icarusNOOPERA}
M.~Antonello {\it et al},
	arXiv:1203.3433

\bibitem{carlipREVIEW} S.~Carlip, Rep.~Prog.~Phys.~64 (2001) 885.

\bibitem{gacQM100} G.~Amelino-Camelia, gr-qc/0012049,
Nature 408 (2000) 661.

\bibitem{cohenglashow}
A.G.~Cohen and S.L.~Glashow ,
Phys.~Rev. Lett.~107 (2011) 181803

\bibitem{bietal}
  X.-J.~Bi, P.-F.~Yin, Z.-H.~Yu and Q.~Yuan,
Phys.~Rev.~Lett.~107  (2011) 241802

\bibitem{cowsiketal}
  R.~Cowsik, S.~Nussinov and U.~Sarkar,
Phys.~Rev.~Lett.~107 (2011) 251801.

\bibitem{cohenglashowV1}
A.G.~Cohen and S.L.~Glashow ,
arXiv:1109.6562v1.

\bibitem{icarusREFUTE}
M. Antonello et al [ICARUS Collaboration], arXiv:1110.3763

\bibitem{noOPERAjune8}
A.~Cho,
{\it Once Again, Physicists Debunk Faster-Than-Light Neutrinos},
ScienceInsider 8 June 2012
(news.sciencemag.org/scienceinsider/)

\bibitem{noglashowLIVma} B.-Q.~Ma,
{\it Mod.~Phys.~Lett.} {\bf 27}, 1230005  (2012).

\bibitem{noglashowLIVgardner}
S.J.~Brodsky and S.~Gardner,
arXiv:1112.1090.

\bibitem{noglashowLIVbezurkov} F.~Bezrukov and H.M.~Lee,
 {\it Phys.~Rev.~D} {\bf 85}, 031901 (2012).

\bibitem{gacdsr1} G.~Amelino-Camelia,
gr-qc/0012051,
{\it Int.~J.~Mod.~Phys.~D} {\bf 11}, 35 (2002).

\bibitem{leedsrPRL}
J.~Magueijo and L.~Smolin,
hep-th/0112090, Phys.~Rev.~Lett.~88 (2002) 190403.

\bibitem{dsrnature} G. Amelino-Camelia,
gr-qc/0207049,
{\it Nature} {\bf 418}, 34 (2002).

\bibitem{jurekDSRnew} J.~Kowalski-Glikman and S.~Nowak,
hep-th/0204245,
Int.~J.~Mod.~Phys.~D12 (2003) 299.

\bibitem{operaDSR}   G.~Amelino-Camelia, L.~Freidel, J.~Kowalski-Glikman and
L.~Smolin,
arXiv:1110.0521.

\bibitem{fransDSROPERA} F.R.~Klinkhamer, arXiv:1110.2146

\bibitem{yiDSROPERA} Y.~Ling,
    arXiv:1111.3716

\bibitem{dimitriDSROPERA} Y.~Huo, T.~Li, Y~.Liao, D.V.~Nanopoulos, Y.~Qi and F.~Wang,
	arXiv:1111.4994

\bibitem{operaLED} S.~Hannestad and M.~Sloth, arXiv:1109.6282.

\bibitem{operaLEDexperts} D.~Marfatia, H.~P\"as, S.~Pakvasa and T.J.~Weiler
arXiv:1112.0527, 	Phys.~Lett.~B707 (2012) 553.

\bibitem{operaSTERILEnew} H.~Minakata and A.Yu.~Smirnov,
arXiv:1202.0953

\bibitem{brustein} R.~Brustein and D.~Semikoz,
	Phys.~Lett.~B706 (2012) 462.

\bibitem{ahlu} D.V.~Ahluwalia, S.P.~Horvath and D.~Schritt,
arXiv:1110.1162

\bibitem{morris}
T.R.~Morris, arXiv:1110.3266

\bibitem{stein2007} A.M.~Steinberg,
Lect.~Notes Phys.~734 (2008) 333

\bibitem{landauerNATURE}
R.~Landauer, Nature 341 (1989) 567.

\bibitem{earlytunnel1} L.A.~MacColl, Phys. Rev. 40 (1932) 621

\bibitem{earlytunnel2} E.P.~Wigner, Phys.~Rev.~98 (1955) 145

\bibitem{bohmbook} D.~Bohm, ``Quantum Theory" (Prentice Hall,
1951)

\bibitem{physrepREVIEW} H.G.~Winful, Phys.~Rep.~436 (2006) 1

\bibitem{recamiREVIEW} V.S.~Olkhovsky and E.~Recami, Phys.~Rep.~214 (1992) 339

\bibitem{steinbergPRL}
A.M.~Steinberg, P.G.~Kwiat and R.Y.~Chiao, Phys.~Rev.~Lett.~71 (1993) 708.

\bibitem{abruptNATURE}
M.D.~Stenner, D.J.~Gauthier and M.A.~Neifeld, Nature 425 (2003)  695.

\bibitem{liparilusignoli} P.~Lipari, M.~Lusignoli and F.~Sartogo,
Phys.~Rev.~Lett.~74 (1995) 4384

\bibitem{pdg} K. Nakamura et al. (Particle Data Group), J.~Phys.~G 37 (2010) 075021.

\bibitem{rickles} D.~Rickles, "Quantum Gravity Meets \& HPS",
  in S. Mauskopf and T. Schmaltz (eds.),
  Integrating History and Philosophy of Science:
  Problems and Prospects, Boston Studies in the
  Philosophy of Science (Springer, Berlin, 2011).

\bibitem{electron1975} Z.G.T.~Guiragossian, G.B.~Rothbart, M.R.~Yearian, R.~Gearhart and J.J.~Murray,
  Phys.\ Rev.\ Lett.\  {34 } (1975)  335.

\end{thebibliography}
\end{document}